# Highlights from the SOAP project survey.
## What Scientists Think about Open Access Publishing
January 20th, 2011

*Suenje Dallmeier-Tiessen [a], Robert Darby [c], Bettina Goerner [b], Jenni Hyppoelae [a], Peter Igo-Kemenes [a,#], Deborah Kahn [d], Simon Lambert [c], Anja Lengenfelder [e], Chris Leonard [c,♮], Salvatore Mele [a,\*], Malgorzata Nowicka [a], Panayiota Polydoratou [e], David Ross [f], Sergio Ruiz-Perez [a], Ralf Schimmer [e], Mark Swaisland [g] and Wim van der Stelt [h]*

[a] CERN, CH1211, Geneva 23, Switzerland
[b] STFC Rutherford Appleton Laboratory, Harwell Science and Innovation Campus, Didcot OX11 0QX, United Kingdom
[c] Springer-Verlag, GmbH, Tiergartenstrasse 17, 69121 Heidelberg, Germany
[d] BioMed Central, 236 Gray's Inn Road, London WC1X 8HL, United Kingdom
[e] Max Planck Digital Library, Amalienstr. 33, 80799 Munich, Germany
[f] SAGE, 1 Oliver's Yard, 55 City Road, London, EC1Y 1SP, United Kingdom
[g] STFC Daresbury Laboratory, Warrington, Cheshire WA4 4AD, United Kingdom
[h] Springer Science+Business Media B.V., Van Godewijckstraat 30, 3311 GX Dordrecht, Netherlands
[#] Also at Gjøvik University College, Po.box 191 Teknologivn. 22, 2802 Gjøvik, Norway
[♮] Now with the Bloomsbury Qatar Foundation Journals, PO Box 582, Doha, Qatar
[*] Corresponding author: Salvatore.Mele@cern.ch

**Abstract**

The SOAP (Study of Open Access Publishing) project has run a large-scale survey of the attitudes of researchers on, and the experiences with, open access publishing. Around forty thousands answers were collected across disciplines and around the world, showing an overwhelming support for the idea of open access, while highlighting funding and (perceived) quality as the main barriers to publishing in open access journals. This article serves as an introduction to the survey and presents this and other highlights from a preliminary analysis of the survey responses. To allow a maximal re-use of the information collected by this survey, the data are hereby released under a CC0 waiver, so to allow libraries, publishers, funding agencies and academics to further analyse risks and opportunities, drivers and barriers, in the transition to open access publishing.

# 1. Introduction

The SOAP (Study of Open Access Publishing) project[1] describes and compares the *offer* and *demand* for open access publishing in peer-reviewed journals. The first phase of the project described the *offer* in open access publishing solutions[2]. In the second phase of the project, the *demand* is assessed through a large-scale survey of scientists across disciplines and around the world, aiming to uncover the attitudes and experiences of scholars with open access publishing. This article timely presents a short summary of the highlights of this survey, whereas a more complete report will follow as the project winds down. This document also serves as an introduction to the survey data, which are hereby released under a CreativeCommons CC0 waiver[3], with the aim of maximizing the scientific return on European Commission research investment by facilitating future academic investigations and by providing small and large publishing enterprises access on equal footing to important market intelligence.

The structure of this article is the following: Section 2 describes the survey structure and dissemination; Section 3 presents highlights on the opinions and attitudes of scholars on open access publishing; Section 4 unveils the barriers reported by survey respondents to adopt open access publishing; Section 5 analyses the experience of researchers who have published in open access journals concerning the possible payment of dedicated fees; Section 6 concludes the document, with additional notes on the release of the survey data in Section 7; Appendix I presents the survey questions.

# 2. The SOAP survey

The SOAP survey was implemented through the popular online SurveyMonkey tool, and comprised a maximum of 23 questions. Appendix I details the entire set of questions, as well as the particular logic applied to skip some questions not relevant for some particular demographics, as identified through responses to earlier questions. Further details are presented in the attached document "SOAP Survey Data – Release Notes", which accompanies the data release and serves as a manual for the data understanding.

The survey was mainly distributed via mailing lists of the publishers participating in the consortium, and in a minor form via members of the OASPA (Open Access Scholarly Publishing Association) and via public mailing lists and newsletters concerned with scholarly communication or specific research fields, as well as targeted mailings to authors in specific scientific communities where response rate was slow or which were not properly covered through other channels[4]. The three largest mailing lists used, and the sources of the largest amount of responses, are, respectively, those of SOAP partners SAGE, Springer and BioMed Central, with 800k, 250k and 170k addresses. The fourth largest mailing was run through Thomson Reuters to 70k authors in fields where, after the first three months of the survey live-time, a relatively low response rate was observed. About 1.5 million individuals are estimated to have been exposed in one way or another to the survey. The survey was "live" for almost seven months, from April 28th, 2010 to November 17th, 2010, even though the vast majority of the responses were collected in the first three-four months. Most of the results presented in this document are based on a snap-shot of the answer to August 10th, 2010, which represents the vast majority of the answers collected in the exercise. Some results, and in particular those representing the answers to open-ended questions, are based on the entire data set. The data released in the attached files corresponds to the entire live-time of the survey.

---

[1] The project is financed by the European Commission under the Seventh Framework Programme, and runs from March 2009 to February 2011. The project is co-ordinated by CERN, the European Organization for Nuclear Research, and is a partnership of publishers (Springer, Sage, BioMed Central), libraries (the Max Planck Digital Library of the Max Planck Society) and funding agencies (the UK Science and Technology Facilities Council). For further information: http://soap-fp7.eu

[2] S. Dallmeier-Tiessen *et al. First results of the SOAP project. Open access publishing in 2010,* http://arxiv.org/abs/1010.0506

[3] http://creativecommons.org/about/cc0

[4] A targeted mailing from the European Commission to around 13'000 European Commission-funded project co-ordinators and alumni of the Marie Curie funding scheme for young researcher collected additional data to an analogous questionnaire. These responses are presented in aggregated form with these results.

Out of 53'890 respondents to the survey to August 10th, 2010, 46'006 identified themselves as active researchers. Out of those, we retained for analysis the answers of 38,358 who published at least one peer-reviewed research article in the last five years and who answered to a key question of the survey about their opinion on open access publishing. The distribution of respondents per high-level discipline is presented in Figure 1[5], responses came from 162 countries, with a large representation from research-intensive nations.

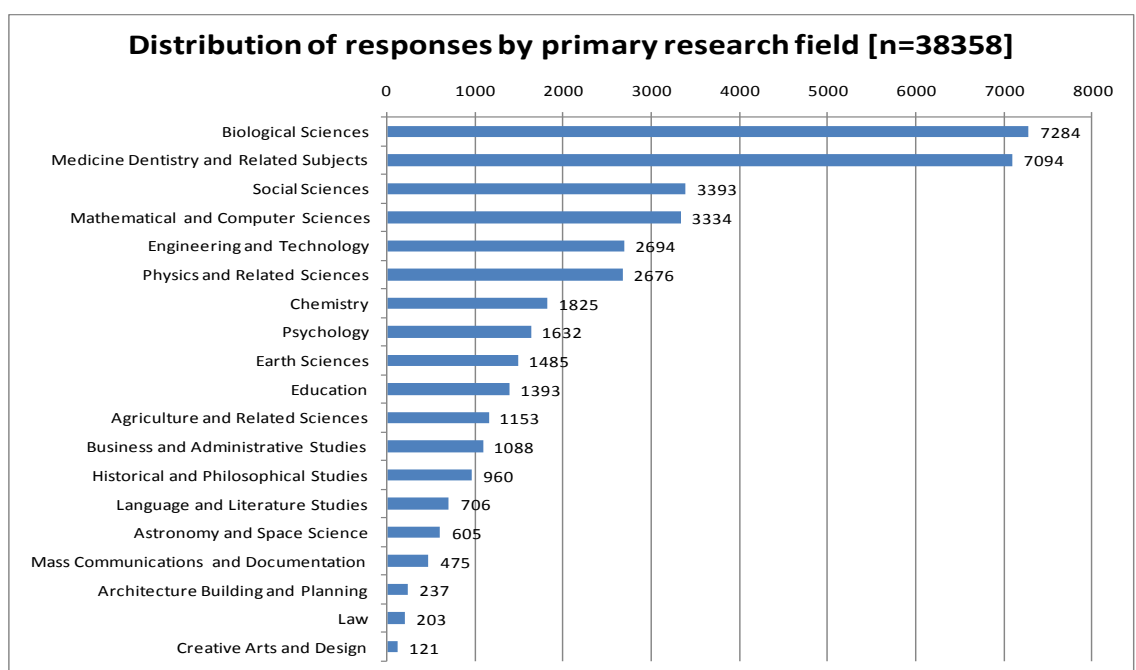

*Figure 1. Distribution of analyzed responses per research field.*

## 3. Attitudes towards open access publishing

One of the key questions asked in the survey is whether respondents considered open access publishing beneficial for their research field. Figure 2 presents the results. In total 89% of published researchers answering to the survey thought that journals publishing open access articles were beneficial for their field. When analysed by discipline, this fraction was higher than 90% in most of the humanities and social sciences, and oscillating around 80% for Chemistry, Astronomy, Physics, Engineering and related disciplines.

The questionnaire allowed respondents to qualify their answers and describe why the considered open access publishing beneficial (or not) in a free-text, open ended, question. Out of the entire survey data set (slightly larger than the one discussed here), 17'852 published researchers answered to this question, contributing a staggering ½ million words on the subject. These answers were all scrutinized and tagged according to one or more recurrent arguments, summarized in Table 1.

---

[5] Respondents could choose a primary research field in a catalogue of 179, as well as adding a second research field. For the sake of simplicity in the presentation of these results, only the primary research field is presented, aggregated at a higher-level taxonomic level.

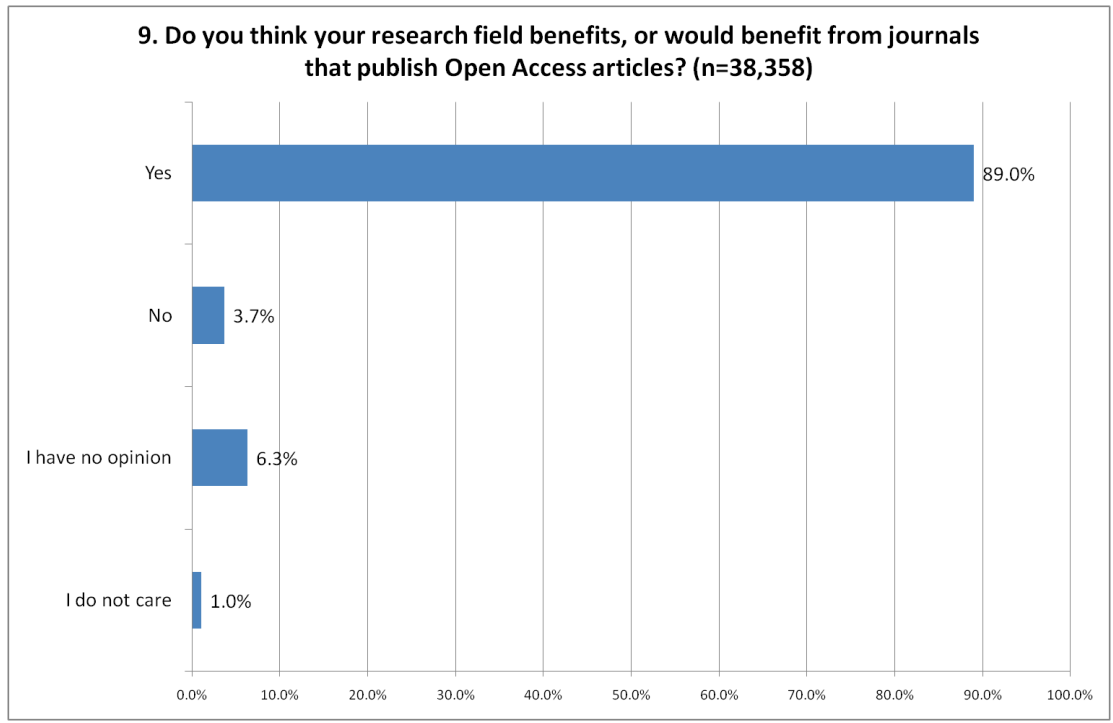

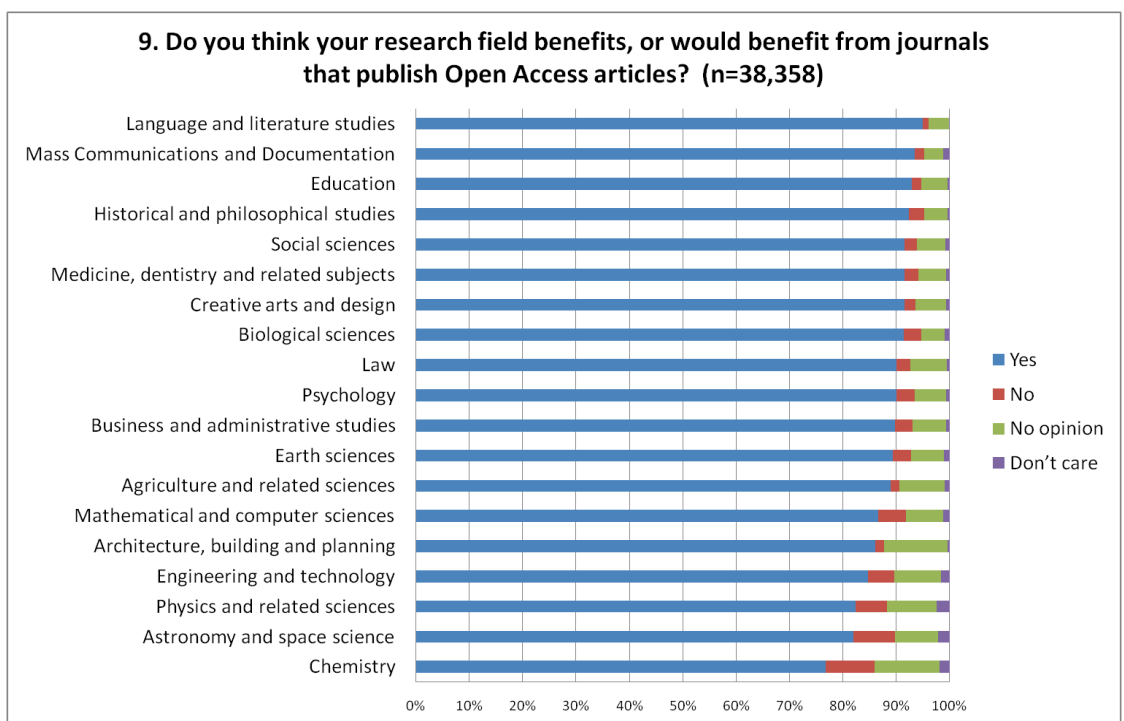

*Figure 2. Fraction of respondents considering open access publishing beneficial for their field in absolute (top) and broken-down by discipline (bottom).*

| |
|---|
| **Accessibility:** refers mostly to technical barriers of accessibility. It has been used for example when a respondent has said that OA would be beneficial as it removes the need to log in on different publisher sites or these can be accessed anywhere, also from home and when travelling. The tag has also been used if the word or concept of 'access' is mentioned but no further explanation is provided, for example if the answer has only been '(because of) ease of access'. |
| **Financial issues:** includes everything related to money: when OA is seen as a better model or solution because of a reason related to financial issues. E.g. 'OA is good because it is free', 'it is cheaper', 'libraries are struggling with current subscription fees', or if there is an idea that a researcher cannot get the information she wants because of lack of individual or library resources. |
| **Individual benefit:** publishing in OA journals is perceived as an asset for an individual researcher to gain more visibility, recognition, readership, citations than the traditional journals. This also includes a saving of time to the individual in the research and publishing process, but does not include the individual benefit a researcher gains when accessing other people's work, what is included in the "scientific community benefit" tag. |
| **Public good:** any benefit to people outside the scientific community. It refers often to moral good, the concept of 'right' or 'fair'. Used for example if developing countries or less privileged entities are mentioned. It is also used for matters of 'principle' e.g. statements as 'all knowledge should be free' or if public funding/tax-payers are mentioned. It also refers to a concept of 'general good' with no other specific reason. |
| **Scientific community benefit** includes all concepts where OA is perceived to benefit the scientific community e.g. by seamless/fast sharing results/methods/information as well as fostering social exchange among researchers. The tag also includes concepts of OA seen as a modern/future/better solution for publishing or when the respondent agrees with OA in principle under condition of quality/peer-review/impact factor comparable or better than traditional or established journals. |
| **Other:** includes all the other goals and ideas. It also includes lack of awareness and other less-frequent concepts. |

*Table 1. Tags used in the analysis of the free-text answers to why respondents consider open access publishing beneficial.*

Figure 3 presents a distribution of the 22'312 tags for the answers of the 16'734 respondents with a positive attitude towards open access publishing. The benefit of open access publishing for the scientific community, including the respondent as reader and scientist, was the most recurring argument at 36%, followed by financial issues and the relevance for the public good, with around 20%. The benefit for the individual, as an author, was fourth at around 10%.

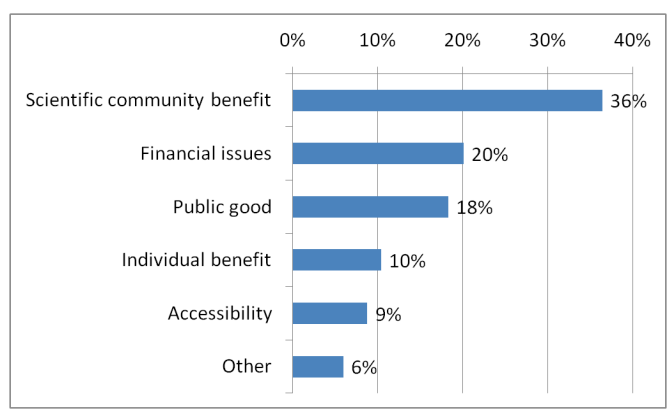

*Figure 3. Distribution of the 22'312 tags for the answers of 16'734 respondents with a positive attitude towards open access publishing.*

An objective of the survey was to assess the degree of agreement of respondents with a series of "myths" about open access publishing. A series of statements, summarized in Table 2, were presented in random order to respondents, who could choose the level at which the agreed with them. Results are presented in a graphical form in Figure 4.

| |
|---|
| Researchers should retain the rights to their published work and allow it to be used by others |
| Open access publishing undermines the system of peer review |
| Open access publishing leads to an increase in the publication of poor quality research |
| If authors pay publication fees to make their articles open access, there will be less money available for research |
| It is not beneficial for the general public to have access to published scientific and medical articles |
| Open access unfairly penalises research-intensive institutions with large publication output by making them pay high costs for publication |
| Publicly-funded research should be made available to be read and used without access barrier |
| Open access publishing is more cost-effective than subscription-based publishing and so will benefit public investment in research |
| Articles that are available by open access are likely to be read and cited more often than those not open access |

*Table 2. Myths about open access publishing with which respondents were asked their level of agreement or disagreement.*

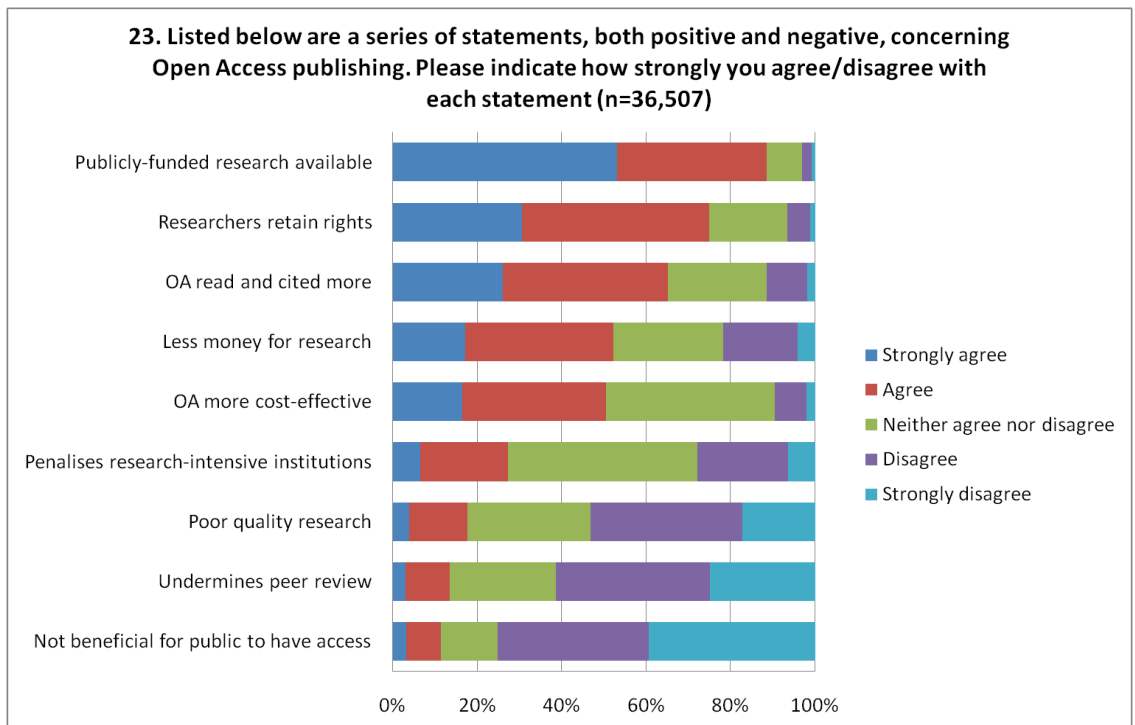

*Figure 4. Fraction of respondents agreeing or disagreeing with a set of statements ("myths") about open access publishing*

## 4. Barriers to open access publishing

Among the respondents to the survey, 29% have not published open access articles. Out of those, 42% admit to having a specific reason not to do so. An open-ended text box was provided for respondents to explain why they had not published open access articles, and 4'976 respondents have contributed their opinion. These answers were all scrutinized and tagged, as for the cases described in Table 1. The most recurring arguments are listed in Table 3.

| |
|---|
| **Accessibility:** the author has had a bad experience with an OA journal, their paper has not been accepted or the respondent thinks there are no OA journals on their field. |
| **Funding:** publication fees or lack of funding for it was mentioned. |
| **Habits:** respondents prefer to publish their papers only in certain established/traditional journals. |
| **Journal quality:** OA journals are perceived/assumed not to be of good quality or they do not have an impact factor. |
| **Next time:** respondents intend to start publishing in OA journals or are already doing so for their next article. |
| **Unawareness:** the respondent is not been aware of OA or OA journals on their field. |
| **Other:** issues such as, but not limited to, the use of green OA to achieve widespread distribution, the inflation of OA journals, the decision taken by other co-authors and other less-frequent concepts. |

*Table 3. Tags used in the analysis of the free-text answers to why were not in a position to publish open access articles.*

Figures 5 present the most recurrent tags for specific reasons not to publish open access articles in general, while Figure 6 presents a breakdown by discipline and by kind of affiliation of the researchers. Overall, the largest barrier is the availability of funding to pay publishing charges, followed by the presence of journals of a (perceived) suitable quality. Across disciplines, the ratio of the barriers change, with funding playing a larger role in fields where long-standing high-quality open access journals are present, and the want of journals playing a larger role where fewer or no open access journals seem to be present. Across different institutions, funding is a larger barrier for over one in two respondents active in hospitals and medical schools.

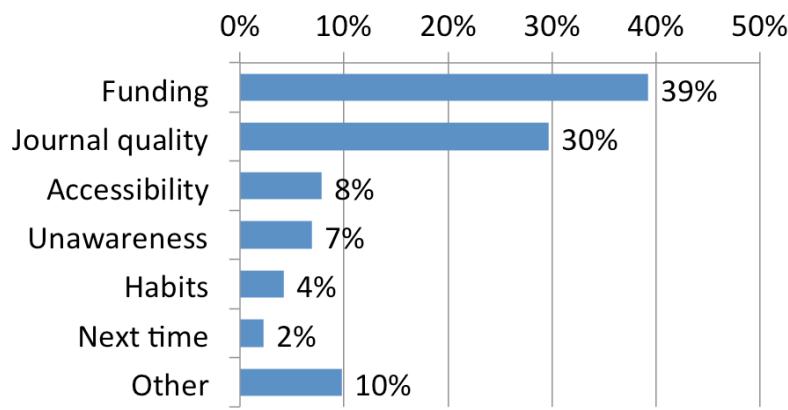

*Figure 5. Specific reasons not to publish open access*

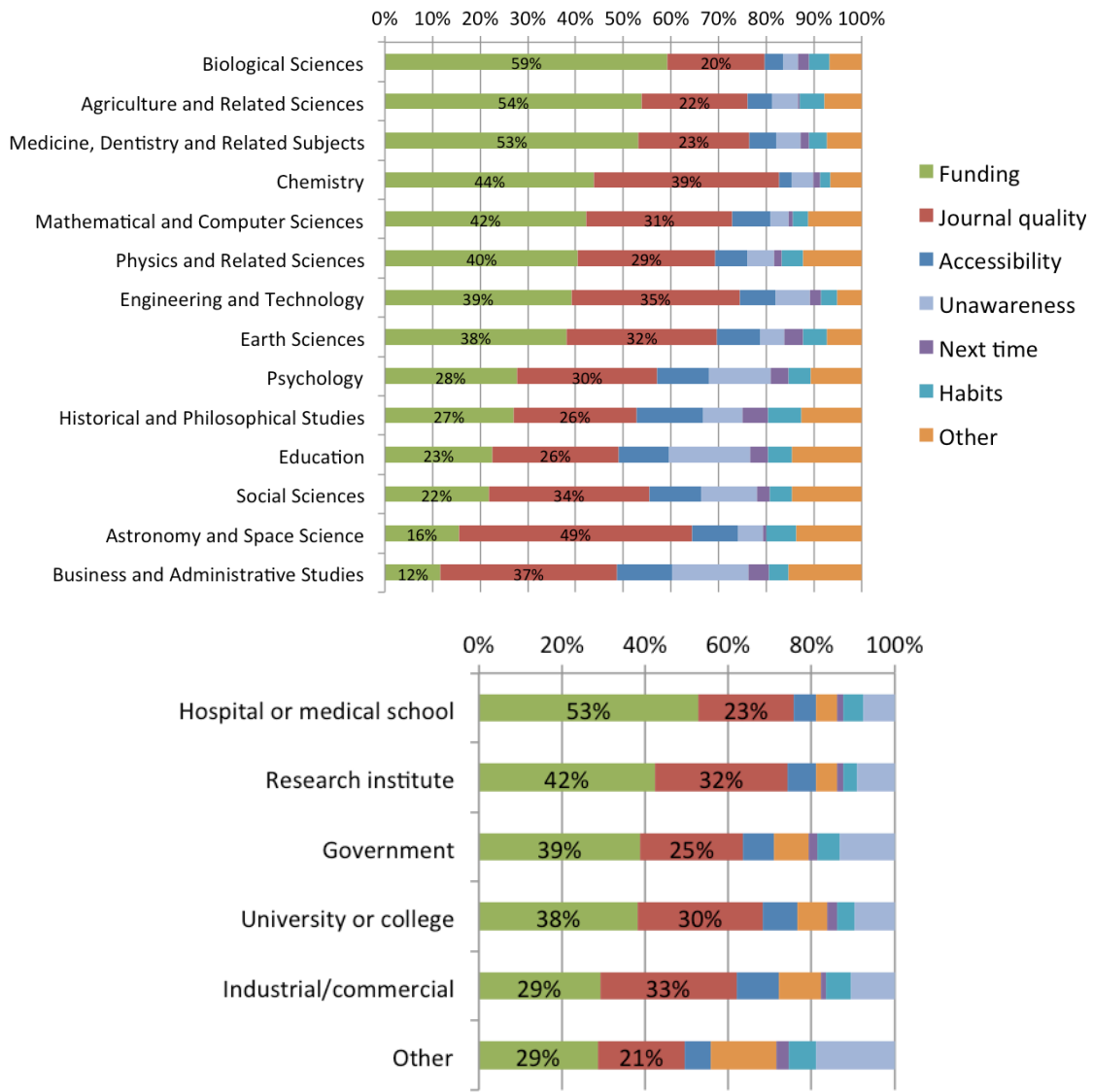

*Figure 6. Specific reasons not to publish open access by discipline (top) and kind of institution (bottom)*

## 5. Experience with open access publishing

Out of the total number of survey respondents, 52% have published at least an open access article, which corresponds to an interesting and unique "survey in the survey" of the experiences of a set of 22'977 scholars who are familiar with this relatively new publishing model. Those respondents were asked several questions out of which a few are singled out for this article and deal with the concept of paying publication fees. A first question, whose answer is graphically depicted in Figure 7 globally and by discipline, concerned the amount of fees paid to publish. Overall, 50% of the respondent had published open access articles without paying a fee, a figure which is much higher for several fields in the humanities and social sciences than in many fields in the natural sciences and engineering.

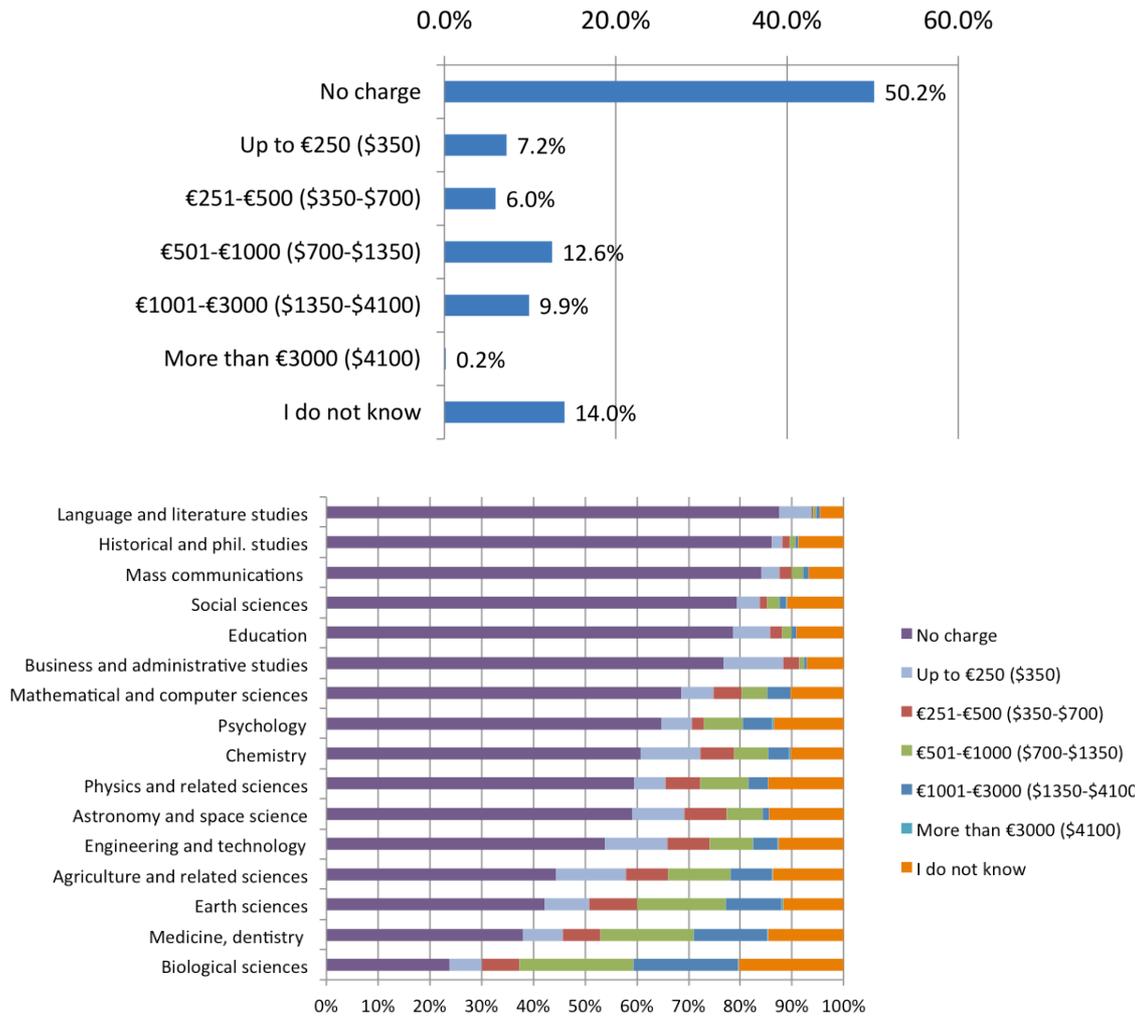

*Figure 7. Publication fee charged for the last open access article published by respondents, overall (top) and by discipline (bottom). 22'977 respondents answered to the question.*

As a follow-up question to the above, respondents were asked how these fees were financed, and the results are presented in Table 4.

| How was this publication fee covered (multiple answers possible) [n=9'645] | |
|---|---|
| My research funding includes money for paying such fees | 28% |
| I used part of my research funding not specifically intended for paying such fees | 31% |
| My institution paid the fees | 24% |
| I paid the costs myself | 12% |
| Other | 5% |

*Table 4. Source of financing for the payment of open access publication fees.*

A follow-up question aimed to clarify how easily funds to pay fees were. Out of the 8'208 respondents to this question, all researchers who had published in open access journals and paid fees to do so, 31% mentioned that finding funds was "easy" and 54% that it was "difficult". The remaining 15% did not use these funds. There are remarkable differences across disciplines and kind of institutions, as presented in Figure 8.

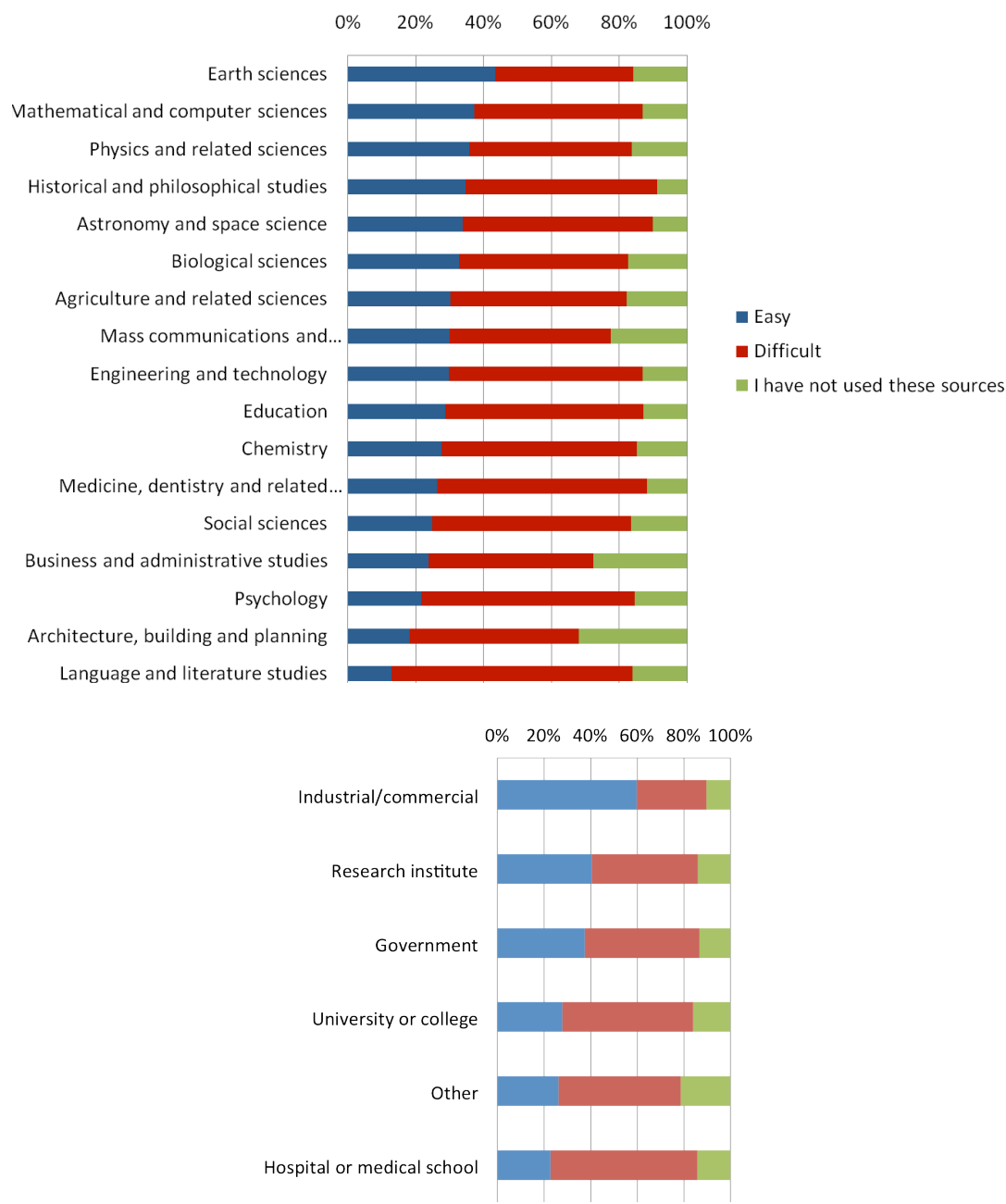

*Figure 8. Ease of access to funds to pay open access publications across disciplines (top) and kind of institutions (bottom)*

## 6. Conclusion

The SOAP survey, the largest to touch issues in open access publishing, has collected a large amount of answer across disciplines and around the world. While the data sample cannot be held to represent the opinions of all scholars active in all countries and in all disciplines, it does present a cross-section of attitudes on open access publishing which were previously not analysed. In addition, a "survey within the survey" of scholars with experience in open access publishing presents novel data on their experience with the process of paying publication fees.

The most relevant findings of the survey are that around 90% of researchers who answered the survey, tens of thousands, are convinced that open access is beneficial for their research field, directly improving the way the scientific community work. At the same time, our previous study[6] found that only 8-10% of articles are published yearly in open access journals. The origin of this gap is apparently mostly due to funding and to the (perceived) lack of high-quality open access journals in particular fields. At the same time, many scientists publish open access articles, without directly incurring costs. Those who do pay fees, however, have a wide varying level of experience on the ease of accessing funds.

This article has only presented some highlights of the SOAP data, more are coming from the project, also in the form of a follow-up survey aimed to clarify questions arising from this preliminary study.

## 7. Survey data release

The project is hereby releasing, in a partially aggregated and filtered form, the survey data, which are hereby released under a CreativeCommons CC0 waiver[7], with the aim of maximizing the scientific return on European Community research investment by facilitating future academic investigations and by providing small and large publishing enterprises access on equal footing to important market intelligence.

The dataset is attached to this article in CSV (comma-separated-values format). MSExcel formats (.xls and .xlsx) are also available from the project website[8]. Release notes, describing the structure and content of the dataset are also included.

We hope that these results and these data could constitute a benchmark against which relate other future academic studies in the field and, at the same time, could inform funding agencies and publishers in their decision concerning the risks and opportunities posed by a transition to open access publishing.

### Acknowledgements
We are indebted to the many individuals and organizations who helped disseminating the survey, as well as the large number of anonymous respondents who took the time to share their opinions and experience about open access publishing. The research results of this project are co-funded by the European Commission under the FP7 Research Infrastructures Grant Agreement Nr. 230220.

---

[6] S. Dallmeier-Tiessen *et al. First results of the SOAP project. Open access publishing in 2010,* http://arxiv.org/abs/1010.0506
[7] http://creativecommons.org/about/cc0
[8] http://soap-fp7.eu

# Appendix I - The survey questionnaire

The entire set of questions asked in the online survey are reproduced in the following. The release notes of the data, in attachment, further describe the treatment of the data.

| |
|---|
| **\*1. Are you involved in research?** <br> I am an active researcher <br> I am in the publishing industry <br> I am a librarian <br> I work in another field and am interested in open access <br><br> [If the answer is anything other than "I am an active researcher", the survey jumps to Q5.] |
| **\*2. Please select your main research field from the drop-down list.** <br>     [Extensive two-level drop-down list of research fields follows] <br><br> **\* Do you wish to include another field of research or add a field that you cannot find in the drop-down list?** <br> Yes <br> No <br>     [If the answer is "Yes", the same list of field is presented for a second choice, plus a text box for "Other"] |
| **\* 3. Which of the following best describes your institution?** <br> University or college <br> Hospital or medical school <br> Research institute <br> Government <br> Industrial/commercial <br> Other |
| **\* 4. How many years have you been employed in research?** <br> Fewer than 5 years <br> 5-14 years <br> 15-24 years <br> 25 years or longer |
| **\* 5. In which country do you work?** <br>     [Drop-down list of countries of the world follows] |
| **6. Please indicate your gender (this question is optional)** <br> Male <br> Female |
| **\* 7. How easily can you gain online access to peer-reviewed journal articles of interest for your research?** <br> Very easily <br> Quite easily <br> With some difficulties <br> I can rarely access the articles I need <br> I do not know |
| Many of the questions that follow concern open access publishing. For the purposes of this survey, an article is open access if its final, peer-reviewed, version is published online by a journal and is free of charge to all users without restrictions on access or use. <br> **\* 8. Do any journals in your research field publish open access articles?** <br> Yes <br> No <br> I do not know |
| **\* 9. Do you think your research field benefits, or would benefit from journals that publish open access articles?** <br> Yes <br> No <br> I have no opinion <br> I do not care |

| Can you briefly explain your opinion? |
| --- |
| [Text box follows] |
| **\* 10. When you are reading a journal article, are you generally aware whether it is open access or not?**<br>Yes<br>No<br>[If the answer is 'No', the survey jumps to Q12.] |
| **\* 11. How do you know whether the article is open access? (Choose more than one answer if applicable)**<br>I had prior knowledge that the article or journal was open access<br>It is clearly indicated on the Web page linking to the article<br>It is clearly indicated in the article itself<br>Other (please specify)<br>    [Text box follows] |
| **\* 12. How many peer reviewed research articles (open access or not open access) have you published in the last five years?**<br>0<br>1-5<br>6-10<br>11-20<br>21-50<br>More than 50<br>[If the answer is "0", the survey jumps to Q20.] |
| **\* 13. What factors are important to you when selecting a journal to publish in?**<br>[Each factor may be rated "Extremely important", "Important", "Less important" or "Irrelevant". The factors are presented in random order.]<br>Importance of the journal for academic promotion, tenure or assessment<br>Recommendation of the journal by my colleagues<br>Positive experience with publisher/editor(s) of the journal<br>The journal is an open access journal<br>Relevance of the journal for my community<br>The journal fits the policy of my organisation<br>Prestige/perceived quality of the journal<br>Likelihood of article acceptance in the journal<br>Absence of journal publication fees (e.g. submission charges, page charges, colour charges)<br>Copyright policy of the journal<br>Journal Impact Factor<br>Speed of publication of the journal<br>Other (please specify)<br>[Text box follows] |
| **\* 14. Who usually decides which journals your articles are submitted to? (Choose more than one answer if applicable)**<br>The decision is my own<br>A collective decision is made with my fellow authors<br>I am advised where to publish by a senior colleague<br>The organisation that finances my research advises me where to publish<br>Other (please specify)<br>    [Text box follows] |
| **\* 15. Approximately how many open access articles have you published in the last five years?**<br>0<br>1-5<br>6-10<br>More than 10<br>I do not know<br>[If the answer is "0", Q16 is asked then the survey jumps to Q20. If the answer is "I do not know", the survey jumps to Q20. Otherwise the survey jumps to Q17.] |
| **\* 16. Has there been a specific reason why you have not published an article by open access? If so, please give your reason(s) in the textbox provided.**<br>Yes<br>No<br><br>Reason(s) for not publishing by open access<br>[Text box follows] |
| **\* 17. What publication fee was charged for the last open access article you published?**<br>No charge |

| | |
|---|---|
| Up to €250 ($350) <br> €251-€500 ($350-$700) <br> €501-€1000 ($700-$1350) <br> €1001-€3000 ($1350-$4100) <br> More than €3000 ($4100) <br> I do not know <br> [If the answer is "No charge" o "I do not know the survey jumps to Q20. | |
| **18. How was this publication fee covered? (Choose more than one answer if applicable)** <br> My research funding includes money for paying such fees <br> I used part of my research funding not specifically intended for paying such fees <br> My institution paid the fees <br> I paid the costs myself <br> Other (please specify) <br>     [Text box follows] | |
| **\* 19. How easy is it to obtain funding if needed for open access publishing from your institution or the organisation mainly responsible for financing your research?** <br> Easy <br> Difficult <br> I have not used these sources | |
| **20. Are you on the editorial board of one or more journals?** <br> Yes <br> No <br> [If the answer is "No", the survey jumps to Q22.] | |
| **21. Are you on the editorial board of any fully open access journals?** <br> Yes <br> No | |
| **22. Do you provide peer review services for one or more journals?** <br> Yes <br> No | |
| **\* 23. Listed below are a series of statements, both positive and negative, concerning open access publishing. Please indicate how strongly you agree/disagree with each statement.** <br> [Each statement may be rated "Strongly agree", "Agree", "Neither agree nor disagree", "Disagree" or "Strongly disagree". The statements are presented in random order.] <br> Researchers should retain the rights to their published work and allow it to be used by others <br> Open access publishing undermines the system of peer review <br> Open access publishing leads to an increase in the publication of poor quality research <br> If authors pay publication fees to make their articles open access, there will be less money available for research <br> It is not beneficial for the general public to have access to published scientific and medical articles <br> open access unfairly penalises research-intensive institutions with large publication output by making them pay high costs for publication <br> Publicly-funded research should be made available to be read and used without access barrier <br> Open access publishing is more cost-effective than subscription-based publishing and so will benefit public investment in research <br> Articles that are available by open access are likely to be read and cited more often than those not open access | |